\documentclass[showpacs,preprintnumbers,amsmath,amssymb,aps,prd,
               nofootinbib,eqsecnum,
	       a4paper]{revtex4}

\def\be{\begin{equation}}
\def\ee{\end{equation}}
\def\bea{\begin{eqnarray}}
\def\eea{\end{eqnarray}}
\def\p{\partial}
\def\n{\nabla}

\def\L{{\pounds}}
\def\R{{{\cal{R}}}}

\def\Q{{\cal{Q}}}
\def\vp{{\varphi}}
\newcommand\dvp[1]{{\delta\varphi_{#1}}}
\newcommand\dvpI[1]{{\delta\varphi_{{#1}I}}}
\newcommand\dvpK[1]{{\delta\varphi_{{#1}K}}}
\newcommand\dvpL[1]{{\delta\varphi_{{#1}L}}}
\newcommand\dU[1]{{\delta U_{#1}}}
\def\S{{\cal{S}}}
\def\H{{\cal H}}
\def\cs2{c_{\rm{s}}^2}
\def\U0{{\bar U_0}}
\def\wt{\widetilde}
\def\dT{{\delta{\bf T}_1}}
\def\dTT{{\delta{\bf T}_2}}
\def\drho{{\delta\rho_1}}
\def\drhorho{{\delta\rho_2}}

\newcommand\eq[1]{Eq.~(\ref{#1})}
\newcommand\eqs[1]{Eqs.~(\ref{#1})}
\begin{document}
\preprint{}
\title{Gauge-invariant perturbations at second order:
 multiple scalar fields on large scales}
\author{Karim A.~Malik}
\affiliation{Cosmology and Astroparticle Physics Group,
Department of Physics, University of Lancaster, Lancaster LA1 4YB,
United Kingdom }
\date{\today}
\begin{abstract}
We derive the governing equations for multiple scalar fields minimally
coupled to gravity in a flat Friedmann-Robertson-Walker (FRW)
background spacetime on large scales. We include scalar perturbations
up to second order and write the equations in terms of physically
transparent gauge-invariant variables at first and second order. This
allows us to write the perturbed Klein-Gordon equation at second order
solely in terms of the field fluctuations on flat slices at first and
second order.
\end{abstract}

\pacs{98.80.Cq \hfill JCAP11 (2005) 005, astro-ph/0506532v3}

\maketitle

%%%%%%%%%%%%%%%%%%%%%%%%%%
\section{Introduction}
%%%%%%%%%%%%%%%%%%%%%%%%%%

Cosmological perturbation theory \cite{Bardeen80,KS} has become the
standard tool to study inflation and its observational consequences
for the Cosmic Microwave Background (CMB) and the formation of large
scale structure \cite{LLBook}.
Linear perturbation theory is sufficient to study the power spectrum
of the primordial perturbations generated during inflation, in
particular to calculate the spectral index and its scale dependence.

One way to glean more information from the CMB is to go beyond first
order perturbation theory, which allows the study of higher order
statistics, such as the bispectrum, or to calculate the amount of
non-gaussianity expected from ones favourite early universe model.

On super-horizon scales, i.e.~scales much larger than the particle
horizon, there are mainly two approaches to study higher order effects
such as non-gaussianity: the first uses second order perturbation
theory following Bardeen
\cite{
Bardeen80,Mukhanov,Bruni,Acquaviva,Maldacena,Nakamura,Noh,Bartolo:2001cw,
Rigopoulos:2002mc,Bernardeau:2002jf,Bernardeau:2002jy,
MW,Bartolo:2004if,Bartolo:2004ty,Enqvist:2004bk,filippo,
Tomita:2005et,Lyth:2005du,Seery:2005wm,Seery:2005gb},
the second uses nonlinear theory and different manifestations of the
separate universe approximation, either employing a gradient expansion, 
as originally used by Salopek and Bond
\cite{SB,RS,Rigopoulos:2004ba,Rigopoulos:2005xx}
or using the $\Delta N$ formalism \cite{SaTa,LMS,Lyth:2005fi}
\footnote{Recently there has also been a covariant study \cite{LV}.}.
In the formalism used by Salopek and Bond \cite{SB} the metric and the
matter variables are not expanded into a power series, instead an
expansion in spatial gradients is used. The $\Delta N$ formalism was
introduced at linear order by Sasaki and Stewart in \cite{SaSt95} (see
also \cite{Starobinsky1982,Starobinsky1986}) and relates the comoving
curvature perturbation to the perturbation in the number of e-foldings
and has recently been extended to the non-linear case
\cite{LMS,Lyth:2005fi}.
In the Bardeen approach the metric and the matter fields are expanded
in a power series in a small parameter.
There has been increased interest in second order perturbation theory
due to the prospect of new high precision data and following the
papers by Acquaviva et al.~\cite{Acquaviva} and Maldacena
\cite{Maldacena} focusing on the study of non-gaussianity from
inflation.

In this article we concentrate on the scalar field dynamics at second
order on large scales in a universe dominated by multiple scalar
fields, including metric perturbations.
In linear theory the Klein-Gordon equation, including metric
perturbations, can be written as a system of coupled evolution
equations in terms of the linear field fluctuations in the flat gauge,
the Sasaki-Mukhanov variables \cite{Sasaki1986,Mukhanov88}.
We show here that the Klein-Gordon equation at second order on large
scales in the multiple field case can also be written solely in terms
of the Sasaki-Mukhanov variables, albeit at second and first order.
The Klein-Gordon equation is linear in the second order perturbations,
with source terms quadratic in the first order perturbations (the same
holds for the field equations).
We also give an expression for the curvature perturbation on uniform
density slices at second order, $\zeta_2$, in terms of the field
fluctuations on flat slices at first and second order.
After calculating the evolution of the field fluctuations at first and
second order we therefore immediately get the evolution of the
curvature perturbation $\zeta_2$. This is similar to the $\Delta N$
formalism, where the evolution of the comoving curvature perturbation
is given by the scalar field dynamics in the background, but which
necessitates the slow roll approximation at higher order.

We consider scalar perturbations including and up to second order in
the metric and in the scalar fields using the Bardeen approach. We
focus on scales larger than the horizon, which allows us to neglect
gradient terms.
We do not include first order vector and tensor perturbations since
they do not couple to the field fluctuations on large scales; also
scalar fields do not support vector modes and the tensor contribution
from inflation is small and constant or decaying. The large scale
focus also makes the equations more transparent and displays the
relevant physics more succinctly, a big bonus at second order. We
leave the inclusion of small scales, and vector and tensor
perturbations to a future publication \cite{M2005}.

The paper is organised as follows: in the next section we give the
field equations and the Klein-Gordon equation up to second order in
gauge dependent form.
In Section \ref{variables_sect}, after reviewing the behaviour of
perturbations at first and second order under gauge transformations,
we construct physically transparent gauge-invariant combinations.
In Section \ref{relate_sect} we show how the gauge-invariant variables
defined in different gauges are related to each other. We also discuss
the construction of total variables in the presence of several fields
or fluids.
We review the derivation of the Klein-Gordon equation at first order
in Section \ref{field_KG1_sect}.
In Section \ref{field2_sect} we give the field equations at second
order in the flat gauge and in Section \ref{KG2_sect} we finally
derive the Klein-Gordon equation at second order. 
We give the second order Klein-Gordon equation in the single field
case in Section \ref{KG2_single}. 
In Section \ref{app_sect} we apply the formalism to a simple two field
inflation model. We conclude in Section \ref{final_sect}.

Throughout this paper we assume a flat Friedmann-Robertson-Walker
(FRW) background spacetime and work in conformal time,
$\eta$. Derivatives with respect to conformal time are denoted by a
dash. Greek indices, $\mu,\nu,\lambda$, run from $0,\ldots3$, while
lower case Latin indices, $i,j,k$, run from $1,\ldots3$.  Upper case
Latin indices, $I,J,K$, denote different scalar fields.

%%%%%%%%%%%%%%%%%%%%%%%%%%%%%
\section{Governing equations}
\label{governing_sect}
%%%%%%%%%%%%%%%%%%%%%%%%%%%%%

The covariant Einstein equations are given by
\be
\label{Einstein}
G_{\mu\nu}=8\pi G \; T_{\mu\nu} \,,
\ee
where $G_{\mu\nu}$ is the Einstein tensor, $T_{\mu\nu}$ the
total energy-momentum tensor, and $G$ Newton's constant.
Through the Bianchi identities, the field equations (\ref{Einstein})
give the local conservation of the total energy and momentum,
\be
\label{nablaTmunu}
\nabla_\mu T^{\mu\nu}=0\,,
\ee
where $\nabla_\mu$ is the covariant derivative.
%
%In the multiple fluid case the total energy-momentum tensor
%is the sum of the energy-momentum tensors of the 
%individual fluids
% 
%\be
%\label{sumTmunu}
%T^{\mu\nu}=\sum_\alpha T^{\mu\nu}_{(\alpha)}\,.
%\ee
%
%
The energy momentum tensor for $N$ scalar fields minimally coupled to
gravity is
\be
\label{multiTmunu}
T_{\mu\nu}=\sum_{K=1}^N\left[
\vp_{K,\mu}\vp_{K,\nu}
-\frac{1}{2}g_{\mu\nu}g^{\alpha\beta}\vp_{K,\alpha}\vp_{K,\beta}\right]
-g_{\mu\nu}U(\vp_1,\ldots,\vp_N)\,, 
\ee
where $\vp_K$ is the $K$th scalar field and $U$ the scalar field
potential and $\vp_{K,\mu}\equiv\frac{\p\vp_{K}}{\p x^\mu}$.\\
%we assumed a Lagrangian of the form
%${\cal{L}}=-U(\vp_K)-\frac{1}{2}\sum$

We split tensorial quantities into a background value and
perturbations according to
\be
\label{tensor_split}
{\bf T}= {\bf T}_0+\dT+\frac{1}{2}\dTT+\ldots\,,
\ee
where the background part is a time dependent quantity only ${\bf
T}_0\equiv {\bf T}_0(\eta)$, whereas the perturbations depend on time
and space coordinates $x^\mu=[\eta,x^i]$, that is ${\bf \delta
T}_n\equiv{\bf \delta T}_n(x^\mu)$. The order of the perturbation is
indicated by a subscript, e.g.~$\dT=O(\epsilon)$. \\

The metric tensor up to second order, including only scalar perturbations, is
\bea
\label{metric1}
g_{00}&=&-a^2\left(1+2\phi_1+\phi_2\right) \,, \\
g_{0i}&=&a^2\left(B_1+\frac{1}{2}B_2\right)_{,i}\,, \\
g_{ij}&=&a^2\left[\left(1-2\psi_1-\psi_2\right)\delta_{ij}
+2E_{1,ij}+E_{2,ij}\right]\,,
\eea
where $a=a(\eta)$ is the scale factor, $\eta$ conformal time,
$\delta_{ij}$ is the flat background metric, $\phi_1$ and $\phi_2$ the
lapse function, and $\psi_1$ and $\psi_2$ the curvature perturbations
at first and second order; $B_1$ and $B_2$ and $E_1$ and $E_2$ are
scalar perturbations describing the shear.
The contravariant form of the metric tensor is given in Appendix
\ref{app_met}.

%%%%%%%%%%%%%%%%%%%%%%%%%%%%%%%%%%%%%%%%
\subsection{Einstein tensor}
\label{ein_tensor_sect}
%%%%%%%%%%%%%%%%%%%%%%%%%%%%%%%%%%%%%%%%

We expand the Einstein tensor in a power series according to
\eq{tensor_split} up to second order
\be
G^\mu_{~\nu}\equiv G^\mu_{(0)\nu} +\delta G^\mu_{(1)\nu}
+\frac{1}{2} \delta G^\mu_{(2)\nu}\,.
\ee
Then the components of the Einstein tensor at zeroth order are
\be
G^0_{~0}=-\frac{3}{a^2}\frac{a'^2}{a^2}\,,
\qquad
G^i_{~j}= \frac{1}{a^2}\left[\frac{a'^2}{a^2}-2\frac{a''}{a}\right]\,,
\ee
at first order
\bea
\label{G00_1}
\delta G^0_{(1)0} &=& 
\frac{6}{a^2}\left[\frac{a'^2}{a^2}\phi_1
+\frac{a'}{a}\psi_1'\right] + O(k^2)\,, \\
\label{G0i_1}
\delta G^0_{(1)i} &=& 
-\frac{1}{a^2}\left[2\psi_1'+2\frac{a'}{a}\phi_1
\right]_{,i} 
\\
\delta G^i_{(1)j} &=& 
\frac{1}{a^2}\left[
4\left(\frac{a''}{a}\phi_1+\frac{a'}{a}\psi_1'\right)
+2\left(\psi_1''+\frac{a'}{a}\phi_1'-\frac{a'^2}{a^2}\phi_1\right)
\right] \delta^i_{~j}+O(k^2)\,,
\eea
and at second order
\bea
\label{G0i_2}
\delta G^0_{(2)0} &=& 
-\frac{6}{a^2}\left[\psi_1'^2-\frac{a'^2}{a^2}\phi_2-\frac{a'}{a}\psi_2'
+4\left(\frac{a'}{a}\phi_1\psi_1'+ \frac{a'^2}{a^2}\phi_1^2 
-\frac{a'}{a}\psi_1\psi_1'\right)
\right] + O(k^2)\,, \\
\label{G00_2}
\delta G^0_{(2)i} &=& -\frac{2}{a^2}\left[
\psi_2'+\frac{a'}{a}\phi_2+4\psi_1\psi_1'-4\frac{a'}{a}\phi_1^2
\right]_{,i} 
-\frac{4}{a^2} \left[2\phi_1\psi'_{1,i}- \psi'_1 \phi_{1,i}\right]  
+O(k^3)\,,
\\
\label{Gij_2}
\delta G^i_{(2)j} &=& \frac{2}{a^2}\left[
{\psi'_1}^2-2\phi_1'\psi_1'
+8\left(\frac{a'}{a}\psi_1\psi_1'-\frac{a'}{a}\phi_1\phi_1'
-\frac{a'}{a}\phi_1\psi_1'-\frac{a''}{a}\phi_1^2\right)
+4\left(\psi_1\psi_1''-\phi_1\psi_1''+\frac{a'^2}{a^2}\phi_1^2\right)
\right.\nonumber\\
&&\qquad\left.
+2\left(\frac{a''}{a}\phi_2+\frac{a'}{a}\psi_2'\right)
+\psi_2''+\frac{a'}{a}\phi_2'-\frac{a'^2}{a^2}\phi_2
\right] \delta^i_{~j}+O(k^2)\,,
\eea
where $O(k^n)$ denotes terms of order $n$ or higher in gradients,
since we are mainly interested in the large results, i.e. the case
$k\to 0$.

%We see that there are no off-diagonal elements at first and second
%order on large scales ($k\ll aH$ or $k\to 0$).

%%%%%%%%%%%%%%%%%%%%%%%%%%%%%%%%%%%%%%%%
\subsection{Energy-momentum tensor}
\label{Tmunu_sect}
%%%%%%%%%%%%%%%%%%%%%%%%%%%%%%%%%%%%%%%%

We split the scalar fields $\vp_I$ into a background and perturbations
up to and including second order according to \eq{tensor_split},
\be
\vp_I(x^\mu)=\vp_{0I}(\eta)+\dvp{1I}(x^\mu)+\frac{1}{2}\dvp{2I}(x^\mu)\,.
\ee
The potential $U\equiv U(\vp_I)$ can be split similarly according to
\be
U(\vp_I)=U_0+\dU1+\frac{1}{2}\dU2\,,
\ee
where
%
%\bea
%\dU1&=&\sum_K\frac{\p U}{\p\vp_K}\dvp{K1}\,,\\
%
%\dU2&=&\sum_K\frac{\p^2 U}{\p\vp_K\p\vp_L}\dvp{K1}\dvp{L1}
%+\sum_K\frac{\p U}{\p\vp_K}\dvp{K2}\,.
%\eea
%
%
\bea
\label{defdU1}
\dU1&=&\sum_K U_{,\vp_K}  \dvp{K1}\,,\\
\label{defdU2}
\dU2&=&\sum_{K,L}U_{,\vp_K\vp_L}\dvp{1K}\dvp{1L}
+\sum_K U_{,\vp_K}\dvp{2K}\,.
\eea
and we use the shorthand $U_{,\vp_K}\equiv\frac{\p U}{\p\vp_K}$.

The energy-momentum tensor for $N$ scalar fields with potential
$U(\vp_I)$ is then split into background, first, and second order
perturbations, using \eq{tensor_split}, as
\be
T^\mu_{~\nu}\equiv T^\mu_{(0)\nu} +\delta T^\mu_{(1)\nu}
+\frac{1}{2} \delta T^\mu_{(2)\nu}\,,
\ee
and we get for the components, from \eq{multiTmunu}, at zeroth order
\bea
T^{0}_{(0)0} &=& -\left(\sum_K\frac{1}{2a^2}{\vp'_{0K}}^2+U_0\right)\,, 
\qquad
T^{i}_{(0)j} =
\left(\frac{1}{2a^2}{\sum_K\vp'_{0K}}^2-U_0\right)\delta^i_{~j}\,,
\eea
at first order
\bea
\label{deltaT100}
\delta T^{0}_{(1)0} &=& -\frac{1}{a^2}\sum_K\left(
\vp_{0K}'\dvpK1'-{\vp'_{0K}}^2\phi_1
\right)-\dU1   \,, \\
\label{deltaT10i}
\delta T^{0}_{(1)i} 
&=& -\frac{1}{a^2}\sum_K\left(\vp_{0K}'\dvpK1_{,i}\right)\,, \\
\delta T^{i}_{(1)j} &=& \frac{1}{a^2}\left[
\sum_K\left(\vp_{0K}'\dvpK1'-{\vp'_{0K}}^2\phi_1\right)-a^2\dU1
\right]\delta^i_{~j}\,,
\eea
and at second order in the perturbations
\bea
\delta T^{0}_{(2)0} &=& -\frac{1}{a^2}\sum_K\left(
\vp_{0K}'\dvpK2'-4\vp_{0K}'\phi_1\dvpK1'-{\vp'_{0K}}^2\phi_2
+4{\vp'_{0K}}^2\phi_1^2+\dvpK1'^2+a^2\dU2\right)+O(k^2)
\,, \\
\label{deltaT20i}
\delta T^{0}_{(2)i} 
&=& -\frac{1}{a^2}\sum_K\left(\vp_{0K}'\dvpK2_{,i}
-4 \phi_1\vp_{0K}'\dvpK1_{,i}+2\dvpK1'\dvpK1_{,i}
\right)\,, \\
\delta T^{i}_{(2)j} 
&=& \frac{1}{a^2}\left[\sum_K\left(
\vp_{0K}'\dvpK2'-4\vp_{0K}'\phi_1\dvpK1'-{\vp'_{0K}}^2\phi_2
+4{\vp'_{0K}}^2\phi_1^2+\dvpK1'^2\right)
-a^2\dU2\right]\delta^i_{~j}+O(k^2) \,.
\eea

The $0-0$ component of the energy momentum tensor gives the
corresponding energy density at zeroth, first and second order for a
universe filled by $N$ scalar fields \cite{Noh}, 
%
%\be
%T^{0}_{~0}\equiv
%-\left(\rho_0+\delta\rho_1+\frac{1}{2}\delta\rho_2
%\right)\,.
%\ee
%
\be 
\label{def_energydensities}
T^{0}_{(0)0} =-\rho_0\,,\qquad\delta T^{0}_{(1)0}
=-\delta\rho_1\,,\qquad\delta T^{0}_{(2)0} =-\delta\rho_2 \,.
\ee

%%%%%%%%%%%%%%%%%%%%%%%%%%%%%%%%%%%%%%%%%
\subsection{Evolution of the background}
\label{background_evol}
%%%%%%%%%%%%%%%%%%%%%%%%%%%%%%%%%%%%%%%%%

The evolution of the background is governed by the Friedmann equation
given from the $0-0$ component of the Einstein equations
\be
\label{friedmann_back}
\H^2=\frac{8\pi G}{3}\left(\frac{1}{2}\sum_K{\vp'_{0K}}^2+a^2U(\vp_{0I})   
\right )\ \,,
\ee
where $\H\equiv\frac{a'}{a}$.  
The scalar fields are governed by the Klein-Gordon equation, which is
from the energy conservation equation, (\ref{nablaTmunu}), at
zeroth order for the $I$th field given by
\bea
\label{KGback}
\vp_{0I}''+2\H\vp_{0I}'+a^2 U_{,\vp_I}=0\,.
\eea
%

%%%%%%%%%%%%%%%%%%%%%%%%%%%%%%%%%%%%%%%%%%%%
\subsection{Perturbed Klein-Gordon equation}
\label{pertKG_sect}
%%%%%%%%%%%%%%%%%%%%%%%%%%%%%%%%%%%%%%%%%%%%

Using the energy momentum tensor given above in Section
\ref{Tmunu_sect} and the energy conservation equation
(\ref{nablaTmunu}) we get the Klein-Gordon equation in the multiple
filed case for the field $I$ at first order
\be
\label{generalKG1}
\dvpI1''+2\H\dvpI1'+2a^2 U_{,\vp_I}\phi_1
-3\vp_{0I}'\psi_1'-\vp_{0I}'\phi'_1+a^2\sum_K U_{,\vp_I\vp_K}\dvpK1
+O(k^2)=0\,.
\ee
We get the Klein-Gordon equation at second order from \eq{nablaTmunu}
and using \eq{generalKG1}
\bea
\label{generalKG2}
\dvpI2''&+&2\H\dvpI2'
+2a^2 U_{,\vp_I}\phi_2-6\psi_1'\dvpI1'-2\phi_1'\dvpI1'
+4\vp_{0I}'\phi_1\phi_1'+4\phi_1a^2\sum_K U_{,\vp_I\vp_K}\dvpK1
\nonumber\\
&-& 12 \vp_{0I}'\psi_1\psi_1'-3\vp_{0I}'\psi_2'-\vp_{0I}'\phi_2'
+a^2\sum_K U_{,\vp_I\vp_K}\dvpK2
+a^2\sum_{K,L} U_{,\vp_I\vp_K\vp_L}\dvpK1\dvpL1
+O(k^2)=0\,.
\eea
Note, that \eqs{generalKG1} and (\ref{generalKG2}) are in an arbitrary
gauge.

%%%%%%%%%%%%%%%%%%%%%%%%%%%%%%%%%%
\section{Variables}
\label{variables_sect}
%%%%%%%%%%%%%%%%%%%%%%%%%%%%%%%%%%

In this section we first calculate the changes coordinate
transformations induce in the metric and matter perturbations at first
and second order. We then construct physically meaningful
gauge-invariant combinations from these variables.

%%%%%%%%%%%%%%%%%%%%%%%%%%%%%%%%%%
\subsection{Gauge transformations}
\label{gauge_trans_sect}
%%%%%%%%%%%%%%%%%%%%%%%%%%%%%%%%%%

We now briefly review how tensorial quantities change under coordinate
transformations \cite{Mukhanov,Bruni}.
We consider two coordinate systems, $\widetilde{x^\mu}$ and ${x^\mu}$,
which are related by the coordinate transformation
\be 
\label{defcoordtrans}
\widetilde{x^\mu} 
= e^{\xi^\lambda \frac{\p}{\p x^\lambda}} x^\mu\,, 
\ee
where $\xi^\lambda$ is the vector field generating the transformation and 
$\xi^\mu\equiv \xi_1^{\mu} +\frac{1}{2}\xi_2^{\mu}+O(\epsilon^3)$.
%where $\xi_1^{\mu}={\cal{O}}(\epsilon)$ and
%$\xi_2^{\mu}={\cal{O}}(\epsilon^2)$.
%
Equation (\ref{defcoordtrans}) can then be expanded up to second-order
as
\bea
\label{coordtrans2}
\widetilde{x^\mu} = x^\mu+\xi_1^{\mu}
+\frac{1}{2}\left(\xi^{\mu}_{1,\nu}\xi_1^{~\nu}+ \xi_2^{\mu}
\right)   \,.
\eea

A tensor ${\bf T}$ transforms under the change of coordinate system
defined above as
\be
\widetilde{\bf T}=e^{\pounds_{\xi^\lambda}}{\bf T}\,,
\ee
where $\pounds_{\xi^\lambda}$ denotes the Lie derivative with respect
to $\xi^\lambda$.
Splitting the tensor$ {\bf T}$ according to \eq{tensor_split} we get
\cite{Mukhanov,Bruni}
\bea
\label{Ttrans}
%\widetilde {\bf T}_0 &=& {\bf T}_0 \nonumber\\
\widetilde \dT
&=& \dT + \L_{\xi_1} {\bf T}_0  \nonumber\\
\widetilde \dTT
&=& \dTT +\L_{\xi_2} {\bf T}_0 +\L^2_{\xi_1}
{\bf T}_0 + 2\L_{\xi_1} \dT
 \,.
\eea

Under a first-order transformation
$\xi_1^\mu=(\alpha_1,\beta_{1,}^{~~i})$, a four scalar such as the
energy density, $\rho$, or the scalar field $\vp$, transforms
therefore as
\be
\label{defrho1}
\widetilde\drho = \drho + \rho_0'\alpha_1 \,,
\ee
and or the first order metric perturbations we have
\bea
\label{defpsi_1}
\widetilde \psi_1 &=& \psi_1-\H\alpha_1 \,,\\
\label{defphi_1}
\widetilde \phi_1 &=& \phi_1+\H\alpha_1+\alpha_1'\,.
\eea

At second order, writing $\xi_2^\mu=(\alpha_2,\beta_{2,}^{~~i})$,
we find  from Eq.~(\ref{Ttrans}) that a four scalar transforms as
\bea
\label{defrho2}
\widetilde\drhorho &=& \drhorho 
+\rho_0'\alpha_2+\alpha_1\left[
\rho_0''\alpha_1+\rho_0'{\alpha_1}'+2\drho'\right]\nonumber\\
&& +\left(2\drho+\rho_0'{\alpha_1}\right)_{,i} \beta_{1,}^{~~i}
\,,
\eea
and the second order metric perturbations transform as
\bea
\label{defpsi_2}
\widetilde \psi_2 &=& \psi_2-\H\alpha_2
-\alpha_1\left[\H{\alpha_1}'
+\left(\H'+2\H^2 \right)\alpha_1
-2\psi_1'-4\H\psi_1\right]\nonumber \\
&& -\left(\H\alpha_1-2\psi_1\right)_{,k}\beta_{1,}^{~~k}\,,\\
\label{defphi_2}
\widetilde \phi_2 &=& \phi_2+\H\alpha_2+{\alpha_2}'
+\alpha_1\left[{\alpha_1}''+5\H{\alpha_1}'
+\left(\H'+2\H^2 \right)\alpha_1
+4\H\phi_1+2\phi_1'\right] \nonumber \\
&&+{\alpha_1}'\left(2{\alpha_1}'+4\phi_1\right)
+\beta_{1,k}\left({\alpha_1}'+\H{\alpha_1}+2\phi_1\right)_{,}^{~k}
+\beta'_{1,k}\left({\alpha_1}-2B_1-\beta_1'\right)_{,}^{~k}
\,. 
\eea

{}From Eqs.~(\ref{defrho2}), (\ref{defpsi_2}), and (\ref{defphi_2}) we
see that on super horizon scales, where gradient terms can be
neglected, the definition of the second order perturbations in the
``new'' coordinate is independent of the spatial coordinate choice
(the ``threading'') at second order in the gradients.
It is therefore sufficient on large scales (at $O(k^2)$) to
specify the time slicing by prescribing $\alpha_1$ and
$\alpha_2$, in order to define gauge-invariant variables
\cite{MW,LMS}.
The procedure to neglect the gradient terms, is explained in
detail in Ref.~\cite{LMS}.
For the approximation to hold one assumes that each quantity can be
treated as smooth on some sufficiently large scale. Formally one
multiplies each spatial gradient $\partial_i$ by a fictitious parameter
$k$, and expands the exact equations as a power series in
$k$, keeping only the zero- and first-order terms, finally
setting $k=1$.

\emph{Note}, in the following sections we shall omit the symbol
$O(k^n)$ denoting the order of the gradient terms neglected, assuming
that if not stated otherwise the equations are valid to $O(k^2)$.

%%%%%%%%%%%%%%%%%%%%%%%%%%%%%%%%%%%%%%%%%%
\subsection{Gauge-invariant combinations}
\label{gi_comb}
%%%%%%%%%%%%%%%%%%%%%%%%%%%%%%%%%%%%%%%%%%

Using the transformation behaviour of the perturbations derived in the
last section we can now construct combinations of these variables that
do not change under gauge transformations, i.e.~gauge-invariant
variables. The combinations given below have definite physical
meaning, such as the curvature perturbation on a hypersurface of
uniform density.

We define the various time slicings and then substitute the results
into the transformation equations to arrive at the gauge-invariant
variables. Whereas the first order results are valid on all scales, we
only consider the large scale case at second order.

%%%%%%%%%%%%%%%%%%%%%%%%%%%
\subsubsection{First order}
%%%%%%%%%%%%%%%%%%%%%%%%%%%

Hypersurfaces of uniform $I$-field, i.e.~$\widetilde{\dvpI1} =0$, are
given from \eq{defrho1} by the temporal slicing
\be
\label{defunifield1}
\alpha_1 =-\frac{\dvpI1}{\vp_{0I}'}\,.
\ee
The curvature perturbation on uniform field hypersurfaces
\cite{Lukash,Lyth1985}, or comoving curvature perturbation, can then be
defined for each field using \eq{defpsi_1} as
\be
\label{defR1I}
\R_{1I}=\psi_1+\H\frac{\dvp1_I}{\vp_{0I}'} \,.
\ee

Flat slices are defined as $\wt{\psi_1}=0$ and we therefore get from
\eq{defpsi_1} the first order time shift
\be
\label{defflat1}
\alpha_1 =\frac{\psi_1}{\H}\,.
\ee
The Sasaki-Mukhanov variable \cite{Sasaki1986,Mukhanov88}, or the
field fluctuation on uniform curvature hypersurfaces, is then given by
\be
\label{defQ1I}
\Q_{1I}\equiv\wt{\dvpI1}=\dvpI1+\frac{\vp_{0I}'}{\H}\psi_1\,.
\ee

The density perturbation on uniform curvature hypersurfaces is defined as
\be
\label{defdeltarho1a}
\wt{\delta\rho_{1\alpha}}=\delta\rho_{1\alpha}
+\frac{\rho_{0\alpha}'}{\H}\psi_1\,,
\ee
where $\rho_{0\alpha}$ and $\delta\rho_{1\alpha}$ are the energy
density of the $\alpha$-fluid in the background and at first order
\footnote{
Greek indices from the beginning of the alphabet,
$\alpha,\beta,\gamma$ will be used to denote different fluids.}.
The curvature perturbation on uniform $\alpha$-fluid density
hypersurfaces is at first order defined as
\be
\label{def_zeta1a}
\zeta_{1\alpha}=-\psi_1-\H\frac{\delta\rho_{1\alpha}}{\rho_{0\alpha}'}\,.
\ee
%
%The curvature perturbation on uniform total density hypersurfaces is
%at first order defined as
%
%\be
%\label{def_zeta1}
%\zeta_1=-\psi_1-\H\frac{\delta\rho_1}{\rho_0'}\,.
%\ee
%

The lapse function on flat slices is from the definition
(\ref{defphi_1}) and \eq{defflat1} given as
\be
\wt{\phi_1}=\phi_1+\frac{1}{\H}\psi_1'
+\left(1-\frac{\H'}{\H^2}\right)\psi_1\,.
\ee

%%%%%%%%%%%%%%%%%%%%%%%%%%%%
\subsubsection{Second order}
%%%%%%%%%%%%%%%%%%%%%%%%%%%%

At second order uniform field slices are, from \eq{defrho2}, defined
by the temporal shift
\be
\label{defunifield2}
\alpha_2 =-\frac{1}{\vp'_{0I}}\left(
\dvpI2-\frac{1}{\vp'_{0I}}\dvpI1'\dvpI1\right)\,,
\ee
on large scales. The curvature perturbation on uniform $I$-field slices
is therefore on large scales at second order, from \eq{defpsi_2} and using 
(\ref{defunifield1}) and (\ref{defunifield2}) \cite{MW}
\be
\label{def_R2I}
\R_{2I}=\psi_2+\frac{\H}{\vp_{0I}'}\dvpI2
-2\frac{\H}{{\vp_{0I}'}^2}\dvpI1'\dvpI1
-2\frac{\dvpI1}{{\vp_{0I}'}}\left(\psi_1'+2\H\psi_1\right)
+\frac{\dvpI1^2}{{\vp_{0I}'}^2}\left(
\H\frac{\vp_{0I}''}{\vp_{0I}}-\H'-2\H^2\right) 
%+O(k^2)
\,.
\ee

The time shift that defines the flat slicing is at second order, from
\eq{defpsi_2}, on large scales given by
\be
\label{defflat2}
\alpha_2 =\frac{1}{\H}\left(
\psi_2+2\psi_1^2+\frac{1}{\H}\psi_1'\psi_1\right)\,.
\ee
Then the field fluctuation on uniform curvature hypersurfaces, or
Sasaki-Mukhanov variable, at second order for the $I$th field, from
\eq{defrho2} and definitions of the time shifts (\ref{defflat1}) and
(\ref{defflat2}) is \cite{MW}
\be
\label{defQ2I}
\Q_{2I}\equiv
\wt{\dvpI2}=\dvpI2+\frac{\vp_{0I}'}{\H}\psi_2
+\left(\frac{\psi_1}{\H}\right)^2\left[
2\H\vp_{0I}'+\vp_{0I}''-\frac{\H'}{\H}\vp_{0I}'\right]
+2\frac{\vp_{0I}'}{\H^2}\psi_1'\psi_1+\frac{2}{\H}\psi_1\dvpI1'
%+O(k^2)
\,.
\ee
The second order lapse function in the flat gauge is given from
Eqs.~(\ref{defphi_2}), (\ref{defflat1}), and (\ref{defflat2}) as
\bea
\wt{\phi_2}&=&\phi_2+\left(1-\frac{\H'}{\H^2}\right)\psi_2
+\frac{1}{\H}\psi_2'
+\frac{2}{\H}\left(5-4\frac{\H'}{\H^2}\right)\psi_1'\psi_1
+\left(
4-6\frac{\H'}{\H^2}-\frac{\H''}{\H^3}+4\frac{\H'^2}{\H^4}
\right)\psi_1^2\nonumber\\
&&+\frac{3}{\H^2}\psi_1'^2+\frac{2}{\H}\psi_1''\psi_1
+4\phi_1\left[
\frac{1}{\H}\psi_1'+\left(1-\frac{\H'}{\H^2}\right)\psi_1\right]
+\frac{2}{\H}\phi_1'\psi_1\,.
\eea

The curvature perturbation on uniform $\alpha$-density hypersurfaces is at
second order on large scales from \eq{defpsi_2} and using 
(\ref{defunifield1}) and (\ref{defunifield2}) given by \cite{MW}
\bea
\label{defzeta2a}
- \zeta_{2\alpha} 
&=&\psi_2+\frac{\H}{\rho_{0\alpha}'}\delta\rho_{2\alpha}
-2\frac{\H}{{\rho_{0\alpha}'}^2}\delta\rho_{1\alpha}'\delta\rho_{1\alpha}
-2\frac{\delta\rho_{1\alpha}}{\rho_{0\alpha}'}\left(\psi_1'+2\H\psi_1\right)
+\frac{{\delta\rho_{1\alpha}}^2}{{\rho_{0\alpha}'}^2}
\left(\H\frac{\rho_{0\alpha}''}{\rho_{0\alpha}'}
-\H'-2\H^2\right) 
\,,
\eea
and from \eq{defrho2} and the definitions of the time shifts
(\ref{defflat1}) and (\ref{defflat2}) we get the second order density
perturbation on uniform curvature hypersurfaces \cite{MW}
\bea
\label{defrho2a}
\widetilde{\delta\rho_{2\alpha}}
&\equiv&\delta\rho_{2\alpha}+\frac{\rho_{0\alpha}'}{\H}\psi_2
+\frac{1}{\H}\left(
2\rho_{0\alpha}'+\frac{\rho_{0\alpha}''}{\H}-\frac{\rho_{0\alpha}'\H'}{\H^2}
\right)\psi_1^2
+2\frac{\rho_{0\alpha}'}{\H^2}\psi_1'\psi_1
+\frac{2}{\H}\psi_1\delta\rho_{1\alpha}'\,.
\eea
%

%%%%%%%%%%%%%%%%%%%%%%%%%%%%%%%%%%%%%%%%%%
\section{Relating variables}
\label{relate_sect}
%%%%%%%%%%%%%%%%%%%%%%%%%%%%%%%%%%%%%%%%%%

In this section we relate the gauge-invariant variables defined above
on different hypersurfaces to each other. Using the field equations we
then define total perturbations from the individual field and fluid
ones, where possible.

%%%%%%%%%%%%%%%%%%%%%%%%%%%
\subsection{First order}
%%%%%%%%%%%%%%%%%%%%%%%%%%%

{}From \eqs{Rvp1} and (\ref{defQ1I}) the Sasaki-Mukhanov variable of the
field $I$ is related to the curvature perturbation on uniform field
hypersurfaces at linear order simply by
\be
\Q_{1I}=\frac{\vp_{0I}'}{\H}\R_{1I}\,.
\ee

{}From \eq{defdeltarho1a} and (\ref{def_zeta1a}) we find that the
curvature perturbation on uniform density hypersurfaces is related to
the density perturbation on uniform curvature hypersurfaces simply as
\be
\zeta_{1\alpha}=-\H\frac{\wt{\delta\rho_{1\alpha}}}{\rho_{0\alpha}'}\,.
\ee

The total density perturbation at first order in the flat gauge is
given in terms of the density perturbations of individual fluids
simply by
\be
\wt{\delta\rho_1}=\sum_\alpha \wt{\delta\rho_{1\alpha}}\,,
\ee
which allows us to relate the total curvature perturbation to the
individual fluid curvature perturbations by
\be
\label{sum_zeta1}
\zeta_{1}=\sum_\alpha \frac{\rho_{0\alpha}'}{\rho_0'}\zeta_{1\alpha}\,.
\ee

We can define the total comoving curvature perturbation (i.e.~relative
to the average fluid velocity or an ``average field'') using the $0-i$
component of the field equations
%energy momentum tensor, \eq{deltaT10i}, and \eq{defR1I}
as
\be
\label{Rvp1}
\R_{1\vp} \equiv \frac{1}{\sum_K{\vp'_{0K}}^2}\sum_I{\vp'_{0I}}^2\R_{1I} \,.
\ee

The total curvature perturbation on uniform density hypersurfaces in
terms of the field fluctuations on flat slices is given by
\be
\label{zeta1sumfield1}
\zeta_1=-\frac{\H}{\sum_L{\vp'_{0L}}^2}
\sum_K{\vp'_{0K}}\wt{\Q_{K1}}\,.
\ee

%%%%%%%%%%%%%%%%%%%%%%%%%%%
\subsection{Second order}
\label{2nd_order_relate_sect}
%%%%%%%%%%%%%%%%%%%%%%%%%%%

The Sasaki-Mukhanov variable at second order defined above in
\eq{defQ2I} can be expressed in terms of the curvature
perturbations on uniform field hypersurfaces, $\R_{1I}$ and $\R_{2I}$,
\be
\label{Q2I_R2I}
\Q_{2I}=\frac{\vp_{0I}'}{\H}\R_{2I}
+\left(\frac{\R_{1I}}{\H}\right)^2\left[
2\H\vp_{0I}'+\vp_{0I}''-\frac{\H'}{\H}\vp_{0I}'\right]
+2\frac{\vp_{0I}'}{\H^2}\R_{1I}'\R_{1I}
%+O(k^2)
\,.
\ee
Similarly we can express the curvature perturbations on uniform field
hypersurfaces in terms of the Sasaki-Mukhanov variables at first and
second order and get, 
\be
\label{R2I_dvpI2}
\R_{2I}=\frac{\H}{\vp_{0I}'}\Q_{2I}
-2\frac{\H}{{\vp_{0I}'}^2}\Q_{1I}'\Q_{1I}
+\frac{\Q_{1I}^2}{{\vp_{0I}'}^2}\left(
\H\frac{\vp_{0I}''}{\vp_{0I}}-\H'-2\H^2\right) 
%+O(k^2)
\,.
\ee
To define the total comoving curvature perturbation at second order
$\R_{2\vp}$ in terms of the $\R_{2I}$ we would need the $0-i$
Einstein equation at this order in an appropriate form, that is
without gradients. Since it is not clear how to arrive at this form of
the $0-i$ Einstein equation (without imposing slow roll, see Section
\ref{field2_sect} below) we shall leave the definition of $\R_{2I}$
open for the moment and shall return to this issue in a future
publication \cite{M2005}.
We note however, that the definition of the total comoving curvature
perturbation at second order $\R_{2\vp}$ is not a problem in itself,
since it was already shown in Ref.~\cite{LMS} that on large scales
$\R_{2\vp}$ and $\zeta_2$ coincide. The definition of the total
curvature perturbation $\zeta_2$ in terms of the field fluctuations is
given below in \eq{zeta2sumfield2}, and we can therefore get
$\R_{2\vp}$ from $\zeta_2$ if we should need it.

The definition of ``total'' quantities from quantities defined for a
specific field or fluid is not more problematic at second order than
at first order, it is a mere question of having the ``right''
equations. As an example we shall now digress slightly from the main
theme of this paper, multiple scalar fields, and analyse the
definition of the total curvature perturbation at second order in the
case of multiple fluids, which is unlike the definition of the total
comoving curvature perturbation, straight forward
\footnote{The relation between scalar fields and fluids at first
order, in particular how fields can be treated as fluids, has been
studied in detail in Ref.~\cite{MW2005}.}.

We can relate the curvature perturbation on uniform
$\alpha$-fluid hypersurfaces, \eq{defzeta2a}, to the density
perturbation on flat slices, \eq{defrho2a},
\be
\label{zeta_rho2}
-\zeta_{2\alpha} =\frac{\H}{\rho_{0\alpha}'}\wt{\delta\rho_{2\alpha}}
-2\frac{\H}{{\rho_{0\alpha}'}^2}\wt{\delta\rho_{1\alpha}}'
\wt{\delta\rho_{1\alpha}}
+\frac{\wt{\delta\rho_{1\alpha}}^2}{{\rho_{0\alpha}'}^2}
\left(\H\frac{\rho_{0\alpha}''}{\rho_{0\alpha}'}
-\H'-2\H^2\right) 
\,,
\ee
and similarly the density perturbation of the $\alpha$-fluid at second
order on flat slices in terms of $\zeta_{1\alpha}$ and
$\zeta_{2\alpha}$,
\be
\label{rho_zeta2}
\wt{\delta\rho_{2\alpha}}
=-\frac{\rho_{0\alpha}'}{\H}\zeta_{2\alpha}
+2\frac{\rho_{0\alpha}'}{\H^2}\zeta_{1\alpha}'\zeta_{1\alpha}
+\zeta_{1\alpha}^2\left(
2\frac{\rho_{0\alpha}'}{\H}+\frac{\rho_{0\alpha}''}{\H^2}
-\frac{\rho_{0\alpha}'\H'}{\H^3}\right)\,.
\ee

The total density perturbation at second order on flat slices is given
in terms of the density perturbations of individual fluids on large
scales simply by
\be
\wt{\delta\rho_2}=\sum_\alpha \wt{\delta\rho_{2\alpha}}\,,
\ee
which allows us to write the total curvature perturbation at second
order to the individual fluid curvature perturbations, using
\eq{rho_zeta2}, as
\be
\label{zeta2sum1}
\zeta_2=\sum_\alpha\frac{\rho_{0\alpha}'}{\rho_0'}\zeta_{2\alpha}
-\left(1+3\cs2\right)\zeta_1^2
+\sum_\alpha\left[
\left(1+3c_\alpha^2\right)\frac{\rho_{0\alpha}'}{\rho_0'}
+\frac{1}{\rho_0'\H}\left(Q_\alpha\frac{\H'}{\H}-Q_\alpha'\right)
\right]\zeta_{1\alpha}^2
-\frac{2}{\H}\sum_\alpha
\zeta_{1\alpha}\left(\zeta_{1\alpha}'-\zeta_{1}'\right)
\,,
\ee
where we used the background energy conservation equation for the
$\alpha$-fluid, \eq{dotrho_a}, given in Appendix \ref{app_back}, 
$c_\alpha^2$ is the adiabatic sound speed of the $\alpha$-fluid, and the
total adiabatic sound speed is denoted $\cs2$.
\footnote{The relation between the total $\zeta_2$ and the curvature
perturbations of the individual fluids, $\zeta_{2\alpha}$, has been
given in the two field case in Ref.~\cite{Bartolo:JCAP}.}

The relative entropy (or isocurvature) perturbation at first order is
defined as \cite{WMLL,MWU}
\be
\label{defS1}
\S_{1\alpha\beta} \equiv 3\left(\zeta_{1\alpha}-\zeta_{1\beta}\right)\,,
\ee
which can be used to rewrite \eq{zeta2sum1} in terms of adiabatic and
non-adiabatic quadratic first order terms as
\bea
\label{zeta2sum2}
\zeta_2&=&\sum_\alpha\frac{\rho_{0\alpha}'}{\rho_0'}\zeta_{2\alpha}
-\frac{2}{3\H}\sum_\alpha\frac{\rho_{0\alpha}'}{\rho_0'}\left(
\zeta_1+\frac{1}{3}\sum_\gamma\frac{\rho_{0\gamma}'}{\rho_0'}\S_{1\alpha\gamma}
\right)
\sum_\gamma\left(\frac{\rho_{0\gamma}'}{\rho_0'}\S_{1\alpha\gamma}\right)'
\nonumber\\
&&+\frac{1}{3}\sum_\alpha\left[
\left(1+3c_\alpha^2\right)\frac{\rho_{0\alpha}'}{\rho_0'}
+\frac{1}{\rho_0'\H}\left(Q_\alpha\frac{\H'}{\H}-Q_\alpha'\right)
\right]
\sum_\gamma\frac{\rho_{0\gamma}'}{\rho_0'}\S_{1\alpha\gamma}
\left(
2\zeta_1
+\frac{1}{3}\sum_\gamma\frac{\rho_{0\gamma}'}{\rho_0'}\S_{1\alpha\gamma}
\right)\,.
\eea
The evolution equation for the relative entropy perturbation at first
order, $\S_{1\alpha\gamma}$, was given in \cite{MW2005}.
We see that there is a purely non-adiabatic contribution to $\zeta_2$,
quadratic in $\S_{1\alpha\gamma}$.
Note that for purely adiabatic first order perturbations
\eq{zeta2sum2} simplifies to
\be
\label{zeta2sumad}
\zeta_2=\sum_\alpha\frac{\rho_{0\alpha}'}{\rho_0'}\zeta_{2\alpha}\,.
\ee

Turning back to scalar fields, the curvature perturbation $\zeta_2$ in
terms of field fluctuations on flat slices at first and second order
is, using \eq{defzeta2a} evaluated for a single fluid, the energy
densities in terms of scalar fields \eq{def_energydensities}, and the
definitions for the Sasaki-Mukhanov variables \eqs{defQ1I} and
(\ref{defQ2I}), given by
%
%V2 sign in front of cs2 corrected
%
\bea
\label{zeta2sumfield2}
\zeta_2&=&\frac{\rho_0}{3 U_0\sum_L{\vp_{0L}'}^2}\sum_K\left(
\vp_{0K}'\Q_{2K}'+{\Q_{1K}'}^2+a^2\wt{\delta U_2}\right)
-\frac{2\H}{\sum_L{\vp_{0L}'}^2}
\left(\frac{2a^2U_0-\sum_L{\vp_{0L}'}^2}{U_0\sum_K{\vp_{0K}'}^2}
\right)\sum_{K,L}U_{,\vp_K}\vp_{0L}'\Q_{1K}\Q_{1L}
\nonumber\\
&&-\H^2\left(\frac{\sum_K\vp_{0K}'\Q_{1K}}{\sum_L{\vp_{0L}'}^2}
\right)^2 
\left[7-3\cs2-\frac{6\sum_L{\vp_{0L}'}^2}{a^2\rho_0}
\right]\,,
\eea
where $\wt{\delta U_2}$ is given from \eq{defdU2} above
evaluated on flat slices, 
\be
\label{defdU2flat}
\wt{\dU2}=\sum_{K,L}U_{,\vp_K\vp_L}\Q_{1K}\Q_{1L}
+\sum_K U_{,\vp_K}\Q_{2K}\,,
\ee
and the adiabatic sound speed for a universe filled by $N$ scalar
fields is given by
\be
\label{cs2field}
\cs2=1+\frac{2}{3\H}
\frac{\sum_K U_{,\vp_K}\vp_{0K}'}{\frac{1}{a^2}\sum_K{\vp_{0K}'}^2}\,,
\ee
and $\rho_0$ is the background energy density defined in
\eq{def_energydensities} above.

Equation~(\ref{zeta2sumfield2}) can be readily evaluated either
analytically or numerically given the Sasaki-Mukhanov variables at
first and second order, $\Q_{1I}$ and $\Q_{2I}$.
However, in Section \ref{KG2_single} below we use $0-i$ Einstein
equation at second order in the single field case and and in Section
\ref{app_sect} we use the slow roll approximation in a particular model
to simplify Eq.~(\ref{zeta2sumfield2}) further, without the time
derivatives of the Sasaki-Mukhanov variables.

%%%%%%%%%%%%%%%%%%%%%%%%%%%%%%%%%%%%%%%%%%%%%%%%%%%%%%%%%%%%%%%%%%%%
\section{Governing equations in the uniform curvature gauge at first 
order}
\label{field_KG1_sect}
%%%%%%%%%%%%%%%%%%%%%%%%%%%%%%%%%%%%%%%%%%%%%%%%%%%%%%%%%%%%%%%%%%%%

We begin this section by giving the field equations at first order in
the uniform curvature gauge and then review the derivation of the
Klein-Gordon equation for multiple scalar fields at first order on
large scales.

%%%%%%%%%%%%%%%%%%%%%%%%%%%%
\subsection{Field equations}
\label{field1_sect}
%%%%%%%%%%%%%%%%%%%%%%%%%%%%

The components of the Einstein tensor and the energy-momentum tensor
are given in Sections \ref{ein_tensor_sect} and \ref{Tmunu_sect}.
Substitution into \eq{Einstein} then gives for the $0-0$ component of
the Einstein equations gives at first order
\be
\label{00Ein1}
2a^2U_0\wt{\phi_1}+\sum_K\vp_{0K}'{\Q_{1K}}'+a^2\wt{\delta U_1}
%+O(k^2)
=0\,,
\ee
and from the $i-j$ component of the Einstein equation we get
\be
\label{ijEin1}
\frac{a'}{a}\wt{\phi_1}'-8\pi G \sum_K\vp_{0K}'{\Q_{1K}}'
%+O(k^2)
=0\,.
\ee

At first order the $0-i$ components of the Einstein tensor,
\eq{G0i_1}, and energy momentum tensor, \eq{deltaT10i}, are linear in
the spatial gradients (by definition the background quantities are
only time dependent), which allows us to write the $0-i$ Einstein
equation without gradients as
\be
\label{0iEin1}
\frac{a'}{a}\wt{\phi_1}-4\pi G \sum_K\vp_{0K}'{\Q_{1K}}=0\,.
\ee

Combining Eqs.~(\ref{00Ein1}) and (\ref{0iEin1}) we get
\be
\label{relation1}
\sum_K\vp_{0K}'{\Q_{1K}}'=-a^2 \sum_K 
\left( U_{,\vp_K} + \frac{8\pi G}{\H}U_0\vp_{0K}'\right)\Q_{1K}\,,
\ee
relating the time derivative of the field fluctuations to the field
fluctuations themselves.

%%%%%%%%%%%%%%%%%%%%%%%%%%%%%%%%%%
\subsection{Klein Gordon equation}
\label{KG1_sect}
%%%%%%%%%%%%%%%%%%%%%%%%%%%%%%%%%%

At first order the Klein-Gordon equation for the field $I$, on flat
slices, is from \eq{generalKG1} given by
\be
\label{KG1_flat}
{\Q_{1I}}''+2\H{\Q_{1I}}'+2a^2 U_{,\vp_I}\wt{\phi_1}
-\vp_{0I}'\wt{\phi_1}'+a^2\sum_K U_{,\vp_I\vp_K}{\Q_{1K}}
%+O(k^2)
=0\,,
\ee
which can be simplified using Eqs.~(\ref{00Ein1}) and (\ref{ijEin1}),
to give
%
%\be
%\wt{\dvpI1}''+2\H\wt{\dvpI1}'
%+\sum_K\left[\left(\frac{\vp_{0I}'}{\H}\right)'
%\frac{\H \vp_{0K}'}{a^2 U_0}\right]\wt{\dvpK1}'
%+\sum_K\left[a^2U_{,\vp_I\vp_K}+\frac{U_{,\vp_K}}{a^2U_0}
%\left(\vp_{0I}'a^2\right)'
%\right]\wt{\dvp1K}
%+O(k^2)=0\,,
%\ee
%
\be
\label{flatKG1_alt}
{\Q_{1I}}''+2\H{\Q_{1I}}'
-\left(\frac{U_{,\vp_I}}{U_0}+\vp_{0I}'\frac{8\pi G}{\H}\right)
\sum_K\vp_{0K}'{\Q_{1K}}'
+a^2\sum_K\left(U_{,\vp_I\vp_K}-\frac{1}{U_0}U_{,\vp_I}U_{,\vp_K}\right)
{\Q_{1K}}
%+O(k^2)
=0\,.
\ee
Equation (\ref{flatKG1_alt}) can be further rewritten using
%Eqs.~(\ref{00Ein1}), (\ref{0iEin1}), or alternatively 
\eq{relation1}
above, to give \cite{Hwang,Taruya,Chris}
\be
\label{flatKG1}
{\Q_{1I}}''+2\H{\Q_{1I}}'
+\sum_K\left[a^2U_{,\vp_I\vp_K}
-\frac{8\pi G}{a^2}\left(a^2\vp_{0I}'\left(
\frac{\vp_{0K}'}{\H}\right)\right)'
\right]{\Q_{1K}}
%+O(k^2)
=0\,,
\ee
which now displays the ``canonical'' time derivatives (in conformal
time) $\p^2/\p\eta^2+2\H \p/\p\eta$, but has a rather involved mass
term.

%%%%%%%%%%%%%%%%%%%%%%%%%%%%%%%%%%%%%%%%%%%%%%%%%%%%%%%%%%%%%%%%%%%%%
\section{Governing equations in the uniform curvature gauge at second 
order}
%%%%%%%%%%%%%%%%%%%%%%%%%%%%%%%%%%%%%%%%%%%%%%%%%%%%%%%%%%%%%%%%%%%%%

In this section we first give the field equation in the uniform
curvature gauge at second order, highlighting problems arising from
the $0-i$ equation at second order and possible ways to solve them.
We then give the Klein-Gordon equation in the multiple field case on
large scales in terms of the Sasaki-Mukhanov variables.

%%%%%%%%%%%%%%%%%%%%%%%%%%%%%%%%%%%%%%%%%%%%%%%%%%%%%%%%%%%%%%%%%
\subsection{Field equations} 
\label{field2_sect}
%%%%%%%%%%%%%%%%%%%%%%%%%%%%%%%%%%%%%%%%%%%%%%%%%%%%%%%%%%%%%%%%%

In this section we present the field equations in the uniform
curvature gauge at second order on large scales.
The components of the Einstein tensor and the energy-momentum tensor
are given in Sections \ref{ein_tensor_sect} and \ref{Tmunu_sect}.
Substitution into \eq{Einstein} then gives for the $0-0$ field
equation in the flat gauge at second order is
\be
\label{00Ein2}
2a^2U_0\left(4\wt{\phi_1}^2-\wt{\phi_2}\right)
=\sum_K\left(
\vp_{0K}'\Q_{2K}'-4\vp_{0K}'\wt{\phi_1}\Q_{1K}'+{\Q_{1K}'}^2\right)
+a^2\wt{\delta U_2} 
%+O(k^2)
\,.
\ee
and, using \eq{0iEin1} and \eq{00Ein2} above, the $i-j$ Einstein
equation can be rewritten as
\be
\label{ijEin2}
\frac{a'}{a}\left(\wt{\phi_2}'-4\wt{\phi_1}'\wt{\phi_1}\right)
=8\pi G\left[\sum_K\left(
\vp_{0K}'\Q_{2K}'+{\Q_{1K}'}^2\right)\right]\,.
\ee

The $0-i$ Einstein equation at second order is given by
\be
\label{0iEin2}
\H\left(\wt{\phi_{2}}_{,i}-4\wt{\phi_1}\wt{\phi_{1}}_{,i}\right)
-4\pi G \sum_K \left(
\vp_{0K}'\Q_{2K,i}+2{\Q_{1K}}'\Q_{1K,i}\right)
+O(k^3)=0\,,
\ee
where we used \eq{0iEin1}. 
The left hand side of \eq{0iEin2} can be rewritten as
$\H\left(
\wt{\phi_2}-2\wt{\phi_1}^2 \right)_{,i}$,
with the spatial derivative {outside} the brackets. However the
right hand side of \eq{0iEin2} can't be written, in the general case, as an
overall gradient, as can be seen above: whereas the $\Q_{2K,i}$ term
is multiplied by a background quantity, similar to the first order
case, the ${\Q_{1K}}'\Q_{1K,i}$ term can't be written as an overall
gradient and the $0-i$ Einstein equation at second order cannot be
brought into scalar form immediately.

In order to recast \eq{0iEin2} in a more useful form, without the
gradients, we have several possibilities:

\begin{itemize}
\item
The easiest solution is to require $\Q_{1K}=const$ or
$\Q_{1K}=0$. Unfortunately this case is not particularly interesting,
since one of the most interesting applications of second order theory
is the study of non-gaussianity, which necessitates the inclusion and
evolution of the first order and in particular the first order squared
terms. We shall therefore not pursue this option further.
\item
The next case is to assume we only have one field to consider and we
shall study this case in detail in Section \ref{KG2_single}. Actually,
we need only one field at first order, but to assume a single field at
first order and multiple field perturbations at second order seems to
be rather contrived.
\item
Another possibility is to use the slow roll approximation such that
\eq{flatKG1} can be rewritten as $\Q_{1I}'\propto f(\vp_{0J})\Q_{1I}$,
where $f(\vp_{0J})$ is a function of the background fields, which
allows as in the single field case to replace $\Q_{1I}'\Q_{1I}$ by
$\Q_{1I}^2$ in \eq{0iEin2} (for a recent detailed exposition of the
slow roll approximation in the multi-field case see
e.g. \cite{Seery:2005gb}).  We shall illustrate this particular case,
i.e.~using the slow roll approximation, by studying a simple two-field
inflation model in Section \ref{app_sect}.
\item
Finally, we can take the divergence of \eq{0iEin2}, which we shall
do next.
\end{itemize}

Following Ref.~\cite{Acquaviva} we take the divergence of \eq{0iEin2} and get
\bea
\label{0iEincontract2}
\H\n^2\wt{\phi_2}&-&4\H\left(\wt{\phi_1}\n^2\wt{\phi_1}
+\wt{\phi_{1}}_{,k}\wt{\phi_{1,}}^{~k}\right)
\nonumber\\
&-&4\pi G \sum_K \left[
\vp_{0K}'\n^2\Q_{2K}+2\Q_{1K}'\n^2\Q_{1K}+
2\Q'_{1K,j}\Q_{1K,}^{~~~~j}\right]+O(k^3)=0\,.
\eea
Using now the inverse Laplacian, defined as $\n^{-2}\n^2 X=X$,
\eq{0iEincontract2} can be rewritten as
\be
\H\left(\wt{\phi_2}-4\wt{\phi_1}^2\right)-4\pi G\sum_K \left[
\vp_{0K}'\Q_{2K}+2\Q_{1K}'\Q_{1K}\right]-R(\wt{\phi_1},\Q_{1I})
+O(k^3)=0\,,
\ee
where we define $R(\wt{\phi_1},\Q_{1I})$ as
\be
R(\wt{\phi_1},\Q_{1I})\equiv 
4\H\left[
\n^{-2}\wt{\phi_1} \n^2\wt{\phi_1}
+\n^{-2}\left(\wt{\phi_{1}}_{,k}\wt{\phi_{1,}}^{~k}\right)
\right]
+8\pi G\sum_K \left[
\n^{-2}\Q_{1K}'\n^{2}\Q_{1K}
+\n^{-2}\left(\Q'_{1K,j}\Q_{1K,}^{~~~~j}\right)
\right]\,.
\ee
But since neither the effect of $R(\wt{\phi_1},\Q_{1I})$ on the
evolution of the field fluctuations nor its large scale limit is
clear, we shall not use this form of the $0-i$ Einstein equation
below.

Not being able to make use of the $0-i$ Einstein equation is
inconvenient, but since it is only a constraint equation and therefore
redundant, we are able to get a closed system of equations without
using it, as shown in Section \ref{KG2_sect} below, which is
sufficient to get the evolution of $\zeta_2$ from
\eq{zeta2sumfield2}. However, the $0-i$ Einstein equation does allow
us to rewrite the Klein Gordon equation in a more compact form with
canonical time derivatives (see above the first order case, Section
\ref{field_KG1_sect}, and below the single field second order case,
Section \ref{KG2_single}).

%%%%%%%%%%%%%%%%%%%%%%%%%%%%%%%%%%%%%%%%%%%%%%%%%%%%%%%
\subsection{Klein Gordon equation} 
\label{KG2_sect}
%%%%%%%%%%%%%%%%%%%%%%%%%%%%%%%%%%%%%%%%%%%%%%%%%%%%%%%

In this section we derive the Klein-Gordon equation in the flat gauge
for $N$ scalar fields on large scales.

%%%%%%%%%%%%%%%%%%%%%%%%%
%\subsubsection{General case}
%\label{KG2_general}
%%%%%%%%%%%%%%%%%%%%%%%%%

At second order the Klein Gordon equations on flat slices is for the
field $I$, from \eq{generalKG2}, on large scales given by
\bea
\label{KG2_flat}
{\Q_{2I}}''&+&2\H{\Q_{2I}}'
+2a^2 U_{,\vp_I}\wt{\phi_2}
-2\wt{\phi_1}'{\Q_{1I}}'
+4\vp_{0I}'\wt{\phi_1}\wt{\phi_1}'
+4\wt{\phi_1}a^2\sum_K U_{,\vp_I\vp_K}{\Q_{1K}}
\nonumber\\
&-& \vp_{0I}'\wt{\phi_2}'
+a^2\sum_K U_{,\vp_I\vp_K}{\Q_{2K}}
+a^2\sum_{K,L} U_{,\vp_I\vp_K\vp_L}{\Q_{1K}}{\Q_{1L}}
%
%+O(k^2)
=0\,.
\eea
We can rewrite \eq{KG2_flat} above, using the field equations at first
and second order given above in Sections \ref{field1_sect} and
\ref{field2_sect}, to substitute for the lapse functions at first and
second order, $\wt{\phi_1}$ and $\wt{\phi_2}$, and get
%
%V2: a^2 corrected
%
\bea
\label{KG2_noncanonic}
{\Q_{2I}}''&+&2\H{\Q_{2I}}'
+a^2\sum_K \left[
U_{,\vp_I\vp_K}-\frac{1}{U_0}U_{,\vp_I}U_{,\vp_K}
\right]{\Q_{2K}}
-\left(\frac{U_{,\vp_I}}{U_0}+\frac{8\pi G}{\H}\vp_{0I}'\right)
\sum_K\left(\vp_{0K}'{\Q_{2K}}'+{\Q_{1K}}'^2\right)\nonumber\\
&+&a^2\sum_{K,L} \left\{
%\left(
U_{,\vp_I\vp_K\vp_L}-\frac{1}{U_0}U_{,\vp_I}U_{,\vp_K\vp_L}
%\right)
%
+\frac{16\pi G}{\H}\sum_{K,L}\left(
U_{,\vp_I\vp_K}-\frac{1}{U_0}U_{,\vp_I}U_{,\vp_K}\right)
\vp_{0L}'\right\}{\Q_{1L}}{\Q_{1K}}\nonumber\\
&+&a^2\frac{16\pi G}{\H}\sum_K
\left(\frac{U_{,\vp_K}}{U_0}+\frac{8\pi G}{\H}\vp_{0K}'\right)
{\Q_{1K}}{\Q_{1I}}'
%+O(k^2)
=0\,.
\eea
This is the gauge-invariant Klein-Gordon equation in the uniform
curvature gauge for $N$ minimally coupled scalar fields on large
scales in terms of the field fluctuations in the flat gauge at first
and second order, $\Q_{1I}$ and $\Q_{2I}$. It is linear in the second
order variables, but quadratic in the first order ones.

%%%%%%%%%%%%%%%%%%%%%%%%%%%%%%%%%%%%%%%%%%%%%%%%%%%%%%%%%
\section{Klein Gordon equation and curvature perturbation:
 single field case}
\label{KG2_single}
%%%%%%%%%%%%%%%%%%%%%%%%%%%%%%%%%%%%%%%%%%%%%%%%%%%%%%%%%

Having dealt with the multi-field case in the previous section, we now
turn to the single field case. The general form of the Klein Gordon
equation at second order, \eq{KG2_noncanonic}, reduces in this case to
\bea
\label{KG2_noncanonic_single}
{\Q_{2}}''&+&2\H{\Q_{2}}'
+a^2 \left[
U_{\vp\vp}-\frac{1}{U_0}U_{\vp}U_{\vp}
\right]{\Q_{2}}
-\left(\frac{U_{\vp}}{U_0}+\frac{8\pi G}{\H}\vp_{0}'\right)
\left(\vp_{0}'{\Q_{2}}'+{\Q_{1}}'^2\right)\nonumber\\
&+&a^2 \left\{
%\left(
U_{,\vp\vp\vp}-\frac{1}{U_0}U_{\vp}U_{\vp\vp}
%\right)
%
+\frac{16\pi G}{\H}\left(
U_{\vp\vp}-\frac{1}{U_0}U_{\vp}U_{\vp}\right)
\vp_{0}'\right\}{\Q_{1}}^2\nonumber\\
&+&a^2\frac{16\pi G}{\H}
\left(\frac{U_{\vp}}{U_0}+\frac{8\pi G}{\H}\vp_{0}'\right)
{\Q_{I}}'{\Q_{K}}
%+O(k^2)
=0\,.
\eea

In the single field case we can use \eq{relation1} to rewrite the
$0-i$ Einstein equation at second order, \eq{0iEin2}, as
\be
\label{0iEin2single}
\H\left(\wt{\phi_{2}}-2\wt{\phi_1}^2\right)_{,i}
=4\pi G \left[
\vp_{0}'\Q_2
-a^2\left(\frac{U_{,\vp}}{\vp_0'}+8\pi G\frac{U_0}{\H}\right)\Q_1^2
\right]_{,i}
+O(k^3)=0\,,
\ee
which allows us to immediately get rid of the spatial gradient in a
similar fashion as in the first order case.
{}From \eqs{00Ein2} and (\ref{0iEin2single}) we then get the useful
relation, similar to \eq{relation1} at first order, 
%
%\bea
%\label{relation2}
%\vp_0'\Q_2'+{\Q_1'}^2&=&-a^2 U_0\left\{
%\left(\frac{U_{,\vp}}{U_0}+\frac{8\pi G}{\H}\vp_0'\right)\Q_2
%\right.\nonumber\\
%%
%&&\left.
%+\frac{U_{,\vp\vp}}{U_0}-{\vp_0'}^2\left(\frac{8\pi G}{\H}\right)^2
%+\frac{8\pi G}{\H}
%\left(\frac{U_{,\vp}}{U_0}+\frac{8\pi G}{\H}\vp_0'\right)
%\left(2\vp_0'-a^2\frac{U_0}{\vp_0'}\right)\Q_1^2
%\right\}\,.
%\eea
%
%
\bea
\label{relation2}
\vp_0'\Q_2'+{\Q_1'}^2&=&-a^2 U_0
\left(\frac{U_{,\vp}}{U_0}+\frac{8\pi G}{\H}\vp_0'\right)\Q_2
-a^2 U_{,\vp\vp}\Q_1^2
\nonumber\\
&&-a^2 U_0\frac{8\pi G}{\H}\left[
2\vp_0'\frac{U_{,\vp}}{U_0}+\frac{8\pi G}{\H}{\vp_0'}^2
-a^2\left(\frac{U_{,\vp}}{\vp_0'}+\frac{8\pi G}{\H}U_0\right)
\right]\Q_1^2 \,.
\eea

Substituting \eq{relation2} into \eq{KG2_noncanonic_single} we arrive at
the single field Klein Gordon equation at second order,
\bea
\label{KG2_canonic_single}
{\Q_{2}}''&+&2\H{\Q_{2}}'
+a^2\left[U_{,\vp\vp}
+\frac{16\pi G}{\H}U_0\vp_0'
\left(\frac{U_{,\vp}}{U_0}+\frac{4\pi G}{\H}\vp_{0}'\right)
\right]\Q_{2}\nonumber\\
&+&a^2 \left\{
U_{,\vp\vp\vp}+\frac{24\pi G}{\H}\vp_0'U_{,\vp\vp}
+\frac{8\pi G}{\H}\left[
3U_{,\vp}\left(\frac{8\pi G}{\H}\left({\vp_0'}^2-2a^2 U_0\right)
-a^2\frac{U_{,\vp}}{\vp_0'}\right)\right.\right.\nonumber\\
&&\qquad\qquad+\left.\left.
\left(\frac{8\pi G}{\H}\right)^2
U_0{\vp_0'}\left({\vp_0'}^2-3a^2 U_0\right)\right]
\right\}\Q_{1}^2
%+O(k^2)
=0\,.
\eea
now with the time derivatives in canonical form, which is nice.

In the single field case the curvature perturbation on uniform density
hypersurfaces at first order in terms of the field fluctuation on the
flat slice is given, from \eq{zeta1sumfield1}, as
\be
\zeta_1=-\frac{\H}{\vp_0'}\Q_1\,,
\ee
and at second order we get from \eq{zeta2sumfield2}, using \eq{relation2}, 
%
%V2: sign corrected
%
\bea
\label{zeta2_single}
\zeta_2=-\frac{\H}{\vp_0'}\Q_2
-\left[7-3\cs2-3\frac{{\vp_0'}^2-a^2U_0}{a^2\rho_0}
+3\frac{a^2 U_{,\vp}}{\H\vp_0'}
\right]\left(\frac{\H}{\vp_0'}\right)^2\Q_1^2
\,.
\eea

Using \eqs{KG2_canonic_single} and (\ref{zeta2_single}) should be
sufficient to show the conservation of $\zeta_2$ on large scales in
the single field case, which was shown in Refs.~\cite{SB,filippo}, and
we will return to this issue in a future work \cite{M2005}.

%%%%%%%%%%%%%%%%%%%%%%%%%%%%%%%%%%%%%%%%%%%%%%
\section{A simple application using slow roll}
\label{app_sect}
%%%%%%%%%%%%%%%%%%%%%%%%%%%%%%%%%%%%%%%%%%%%%%

We can now apply the formalism derived in the previous sections to the
simple two field model of Ref.~\cite{Enqvist:2004bk} which has
subsequently been studied in Refs.~\cite{Lyth:2005du,Lyth:2005fi} and
most recently in Ref.~\cite{antti} (for earlier work on two-field
inflation at linear order see e.g.~\cite{Polarski:1992dq}). Using the
slow-roll approximation we calculate the curvature perturbation at
second order, $\zeta_2$, in terms of the field fluctuations.

The potential is given as
\be
\label{potential}
U=\U0+\frac{1}{2}m_1^2\vp_1^2+\frac{1}{2}m_2^2\vp_2^2\,,
\ee
with the first term, $\U0=const$, dominating and where $m_I$ denotes
the mass of the $I$th field. Furthermore we assume slow roll and
\be
\label{condition}
\vp_{02}=0\,,\qquad\vp_{02}'=0\,.
\ee
The background Friedmann constraint, \eq{friedmann_back}, then simplifies to
\be
\H^2=\frac{8\pi G}{3}\U0 \,,
\ee
and the background Klein-Gordon equation, \eq{KGback}, gives
\be
\vp_{01}'+\frac{a^2 m_1^2}{3\H}\vp_{01}=0\,.
\ee

{}From \eq{zeta1sumfield1} we immediately get the curvature
perturbation at first order in terms of the field fluctuations on flat
slices in this model,
\be
\zeta_1=8\pi G\frac{\U0}{m_1^2\vp_{01}}\Q_{11}\,.
\ee

The calculation of the second order curvature perturbation in terms of
the field fluctuations is slightly more involved. As pointed out in
Section \ref{2nd_order_relate_sect}, in order to get $\zeta_2$ solely
in terms of the field fluctuations without their time derivatives we
have to use the $0-i$ Einstein equation at second order, and, as
discussed in Section \ref{field2_sect}, to get the latter in a useful
form we need the first order Klein-Gordon equations.

The perturbed Klein-Gordon equation at first order on flat slices,
\eq{KG1_flat}, reduces in the slow roll limit for the model specified
by \eqs{potential} and (\ref{condition}) to
\be
\label{KG1_slowroll}
\Q_{11}'+\frac{a^2 m_1^2}{3\H}\Q_{11}=0\,, \qquad
\Q_{12}'+\frac{a^2 m_2^2}{3\H}\Q_{12}=0\,,
\ee
where $\Q_{11}$ and $\Q_{12}$ are the field fluctuations in the flat
gauge, the Sasaki-Mukhanov variables, at first order for the two
fields.

The $0-i$ equation at second order on flat slices, Eq.~(\ref{0iEin2}),
simplifies then using \eqs{KG1_slowroll} to
\be
\H\left(\wt\phi_2-2\wt\phi_1^2\right)
=4\pi G\left[
\vp_{01}'\Q_{21}-\frac{a^2 m_1^2}{3\H}\Q_{11}^2
-\frac{a^2 m_2^2}{3\H}\Q_{12}^2\right]\,.
\ee
Rewriting \eq{00Ein2}, the $0-0$ Einstein equation at second order, in
a similar fashion we arrive at the useful relation
\be
\sum_K\left(\vp_{0K}\Q_{2K}'+{\Q_{1K}'}^2\right)+a^2\delta U_2
=a^2\left(m_1^2\vp_{01}\Q_{21}+m_1^2\Q_{11}^2+m_2^2\Q_{12}^2\right)\,,
\ee
which upon substitution in \eq{zeta2sumfield2} yields
\be
\label{zeta2_model}
\zeta_2=8\pi G\frac{\U0}{m_1^2\vp_{01}^2}\left[
\vp_{01}\Q_{21}+\frac{m_2^2}{m_1^2}\Q_{12}^2
+\left(1+2\frac{8\pi G\U0}{m_1^2}\right)\Q_{11}^2
\right]\,,
\ee
the curvature perturbation on uniform density hypersurfaces at second
order in terms of the field fluctuations on flat slices for the model
specified above.

This expression for $\zeta_2$, Eq.~(\ref{zeta2_model}) above, agrees
with the one found by Lyth and Rodriguez in Ref.~\cite{Lyth:2005fi}
using the $\Delta N$ formalism, if we take into account that
$\zeta_2=\zeta_{2\rm{LR}}+2{\zeta_1}^2$ \cite{LMS}. In particular we
find that $\zeta_2$ does not contain any non-local terms.

However, the curvature perturbation $\zeta_2$ found here disagrees
with the expression for the second order curvature perturbation found
by Enqvist and Vaihkonen in Ref.~\cite{Enqvist:2004bk}, although the
order of magnitude estimate published subsequently by Vaihkonen
\cite{antti} seems to agree with the result found here.

%%%%%%%%%%%%%%%%%%%%%%%%%%%%%%%%%%%%
\section{Discussion and conclusions}
\label{final_sect}
%%%%%%%%%%%%%%%%%%%%%%%%%%%%%%%%%%%%

In this paper we have presented the Klein-Gordon equation for multiple
minimally coupled scalar fields at second order in the perturbations
in a perturbed FRW background on large scales. 
We have shown that using suitable gauge-invariant variables, namely
the field fluctuations in the flat gauge or Sasaki-Mukhanov variables,
the Klein-Gordon equation at second order, \eq{KG2_noncanonic}, can be
written solely in terms of these variables, as in the first order
case, and is linear in the second order variables, but has source
terms quadratic in the first order field fluctuations.

We have also given the relation between gauge-invariant quantities in
different gauges, and hence on different hypersurfaces, which at second
order is non-trivial. In particular we give the curvature perturbation
on uniform density hypersurfaces, $\zeta_2$, in terms of the
Sasaki-Mukhanov variables of the individual fields, $\Q_{1I}$ and
$\Q_{2I}$ in \eq{zeta2sumfield2}.
We calculated $\zeta_2$ for a particular two-field model during
slow-roll inflation, \eq{zeta2_model}, and found excellent agreement
with the expression derived using the $\Delta N$ formalism
\cite{Lyth:2005fi}. In particular we find using second order
cosmological perturbation theory that there are no non-local terms in
the expression for $\zeta_2$ in this model.

Having an expression for $\zeta_2$ solely in terms of the field
fluctuations allows us to get the evolution of $\zeta_2$ directly
from solving the Klein-Gordon equations at zeroth, first, and second
order, \eqs{KGback}, (\ref{flatKG1}), and (\ref{KG2_noncanonic}),
respectively, together with the Friedmann constraint,
\eq{friedmann_back}. Alternatively we could have calculated the
non-adiabatic pressure due to the multiple scalar fields first and
then solve an evolution equation for $\zeta_2$ \cite{WMLL,MW,MW2005},
but using \eq{zeta2sumfield2} eliminates the integration of the
$\zeta_2$ evolution equation and is therefore simpler, since the
Klein-Gordon equations have to be solved in both cases.

Given suitable initial conditions for the Klein-Gordon equations
governing the dynamics of the fields at zeroth, first, and second
order the equations can be integrated numerically.
The equations can be solved order by order, the background and first
order fields acting as source terms for the second order equations,
since we do not consider back-reaction of the higher order
perturbations on the lower order ones. Solving the second order
equations numerically will therefore be very similar to solving the
first order system.

Alternatively, employing a slow roll approximation at all orders, the
equations can be approximated analytically.
It will be particularly interesting to compare the evolution of
$\zeta_2$ from a slow roll version of the Klein-Gordon equation,
\eq{KG2_noncanonic}, to the $\Delta N$-formalism \cite{Lyth:2005fi},
which also uses slow roll and also gives the evolution of the
curvature perturbation directly from the solution of the Klein-Gordon
equations, without having to integrate an evolution equation for
$\zeta_2$.

The next step will be the extension of the formalism presented here to
small scales, i.e.~deriving a formalism valid on all scales. This will
also be an advantage compared to other approaches based on the
separate universe paradigm, such as the $\Delta N$-formalism, which
can not be extended to small scales since it is by assumption only
valid on scales of order or larger than the horizon.
First steps towards extending the formalism are under way at present,
based on the work presented here \cite{M2005}.

%%%%%%%%%%%%%%%%
\acknowledgments
%%%%%%%%%%%%%%%%

The author is grateful to David Lyth, David Matravers, and David Wands
for useful discussions and comments. KAM is supported by PPARC grant
PPA/G/S/2002/00098.
Algebraic computations of tensor components were performed using
the \textsc{GRTensorII} package for Maple.

%%%%%%%%%%%%
\appendix
%%%%%%%%%%%%

%%%%%%%%%%%%%%%%%%
%\section{This and that}
%%%%%%%%%%%%%%%%%%

%%%%%%%%%%%%%%%%%%%%%%%%%%%%%%%%%%%%%%%%%%%
\section{Background field equations}
\label{app_back}
%%%%%%%%%%%%%%%%%%%%%%%%%%%%%%%%%%%%%%%%%%%

The Friedmann equation is given from the $0-0$ component of the
Einstein equations
\be
\H^2=\frac{8\pi G}{3}a^2\rho_0 \,,
\ee
where $\H\equiv\frac{a'}{a}$. From the $i-j$ component we find
\be
\label{ij_back}
\left(\frac{a'}{a}\right)^2-2\frac{a''}{a}=8\pi G a^2 P_0\,.
\ee
The two previous equations can be rewritten as
\be
\H'=-\frac{4\pi G}{3}a^2\left(\rho_0+3P_0\right)\,,
\ee
or alternatively as
\be
\frac{a''}{a}=\frac{4\pi G}{3}a^2\left(\rho_0-3P_0\right)\,.
\ee
Also useful are
\be
2\left(\frac{a'}{a}\right)^2-\frac{a''}{a}
=4\pi G a^2 \left(P_0+\rho_0\right)\,,
\ee
where the right hand side in the case of scalar fields simplifies to 
%
%\be
$a^2\left(P_0+\rho_0\right)=\sum_K{\vp'_{0K}}^2$, and
%\ee
%
%Finally,
%
\be
\H'+2\H^2
=4\pi G a^2 \left(\rho_0-P_0\right)\,,
\ee
where the right hand side in the case of scalar fields simplifies to 
%
%\be
$\left(\rho_0-P_0\right)=2U_0$.\\
%\ee

The background energy conservation for the $\alpha$-fluid is given by
\cite{KS,MW2005}
\be
\label{dotrho_a}
\rho_{0\alpha}'=-3\H\left(\rho_{0\alpha}+P_{0\alpha} \right)+Q_\alpha\,,
\ee
where $\rho_{0\alpha}$, $P_{0\alpha}$, and $Q_\alpha$ are the energy
density, the pressure and the energy transfer to the
$\alpha$-fluid. Note that the energy transfer defined here is related
to the one define in \cite{MW2005}, $\hat Q_\alpha$, by
$aQ_\alpha=\hat Q_\alpha$.
The adiabatic sound speed of the $\alpha$-fluid is
\be
\label{c2a}
c_\alpha^2\equiv \frac{P_{0\alpha}'}{\rho_{0\alpha}'}\,,
\ee
related to the total adiabatic sound speed by 
\be
\label{def_cs2}
\cs2=\sum_\alpha \frac{\rho_{0\alpha}'}{\rho_0'}c_\alpha^2
\,.
\ee
For more details on the multi-fluid formalism see \cite{MW2005}.

%%%%%%%%%%%%%%%%%%%%%%%%%%%%%%%%%%%%%%%%%%%
\section{The metric tensor}
\label{app_met}
%%%%%%%%%%%%%%%%%%%%%%%%%%%%%%%%%%%%%%%%%%%

The metric tensor up to second order, including only scalar perturbations, is
\bea
%\label{metric1}
%
g_{00}&=&-a^2\left(1+2\phi_1+\phi_2\right) \,, \\
g_{0i}&=&a^2\left(B_1+\frac{1}{2}B_2\right)_{,i}\,, \\
g_{ij}&=&a^2\left[\left(1-2\psi_1-\psi_2\right)\delta_{ij}
+2E_{1,ij}+E_{2,ij}\right]\,.
\eea
and its contravariant form is
\bea
%\label{metric2}
%
g^{00}&=&-a^{-2}\left[1-2\phi_1-\phi_2+4\phi_1^2-
B_{1,k}B_{1,}^{~k}\right] \,, \\
g^{0i}&=&a^{-2}\left[B_{1,}^{~i}+\frac{1}{2}B_{2,}^{~i}
-2B_{1,k}E_{1,}^{~ki}+2\left(\psi_1-\phi_1\right)B_{1,}^{~i}
\right]\,, \\
g^{ij}&=&a^{-2}\left[\left(1+2\psi_1+\psi_2+4\psi_1^2\right)\delta^{ij}
-\left(
2E_{1,}^{~ij}+E_{2,}^{~ij}-4E_{1,}^{~ik}E_{1,k}^{~~j}
+8\psi_1E_{1,}^{~ij}+B_{1,}^{~i}B_{1,}^{~j}\right)
\right]\,,
\eea
%

%%%%%%%%%%%%%%%%%%%%%%%%%%%%
{}
%%%%%%%%%%%%%%%%%%%%%%%%%%

%%%%%%%%%%%%%%%%%%%%%%%%%%%%%%%
\end{document}